\documentclass[11pt, a4paper,twocolumn]{article}

\usepackage{graphicx}
\usepackage{amsmath}
\usepackage{amsfonts}
\usepackage{amssymb}
\usepackage{hyperref}
\usepackage{slashed}

\newcommand{\be}{\begin{equation}}
\newcommand{\ee}{\end{equation}}
\newcommand{\bea}{\begin{eqnarray}}
\newcommand{\eea}{\end{eqnarray}}
\newcommand{\ba}{\begin{array}{ll}}
\newcommand{\ea}{\end{array}}

\renewcommand\a{\alpha}

\newcommand\g{\gamma}

\renewcommand\P{\ensuremath{\mathcal P}}
\newcommand\C{\ensuremath{\mathcal C}}

\newcommand\eV{\text{eV}}

\newcommand\MeV{\text{MeV}}
\newcommand\GeV{\text{GeV}}
\newcommand\TeV{\text{TeV}}
\newcommand\meV{\text{meV}}

\newcommand\BB{\ensuremath{0\nu2\beta}}

\begin{document}

\twocolumn[
  \begin{@twocolumnfalse}
\def\lsim{\mathrel{\raise.3ex\hbox{$<$\kern-.75em\lower1ex\hbox{$\sim$}}}}
\def\gsim{\mathrel{\raise.3ex\hbox{$>$\kern-.75em\lower1ex\hbox{$\sim$}}}}
\sf
\centerline{\Huge  Seesaw at LHC through Left - Right Symmetry$^{\S}$  }
\vspace{5mm}
\centerline{\large Goran Senjanovi\' c}
\centerline{{\it ICTP, Trieste, Italy }}
\vspace{5mm}
\centerline{\large\sc Abstract}
\begin{quote}
\small
 
   I argue that LHC may shed light on the nature of neutrino mass through the 
   probe of the seesaw mechanism.  The smoking gun signature is lepton number
   violation through the production of same sign lepton pairs, a collider analogy of
   the neutrinoless double beta decay.
 I discuss this in the context of $L-R$ symmetric theories, which led originally to neutrino mass and the seesaw mechanism.  A $W_R$ gauge boson with a mass in a few TeV region could 
easily dominate neutrinoless double beta decay, and its discovery at LHC would have spectacular signatures
of parity restoration and lepton number violation. Moreover, LHC can measure the masses of the right-handed 
neutrinos and the right-handed leptonic mixing matrix, which could in turn be used to predict the rates for 
neutrinoless double decay and lepton flavor violating violating processes. The LR scale at the LHC energies offers great hope
of observing these low energy processes in the present and upcoming experiments.

\end{quote}
  \end{@twocolumnfalse}]
 {
 \renewcommand{\thefootnote}%
   {\fnsymbol{footnote}}
 \footnotetext[4]{Based on the plenary talk at the Conference in Honour of Murray Gell-MannÕs 80th Birthday, 
 NTU, Singapore, February 2010. New results, after the conference, are  taken from a recent study by Tello, Nemev\v{s}ek, Nesti, Vissani and the author,  Ref. \cite{Tello:2010am}.}
}

 \rm
\newpage
\tableofcontents

 \section{Foreword} 
  
    It gives me both great pleasure and honour to make part of the celebration of Murray Gell-MannÕs 80th birthday.
  I first came across of his name in 1965 at the tender age of fifteen when I heard of mysterious quarks. I would keep
  hearing his name continuously in the years to follow, but 
 my first physical contact with Gell-Mann came some fourteen years later when he gave a summary talk at a big conference at
    Caltech. In his talk Gell-Mann not only mentioned my work with Rabi Mohapatra on the strong CP problem 
    but also perfectly pronounced my last name, the first time ever in the USA. I felt rather proud and only later I noticed that he 
    pronounced perfectly every name, Italian, Chinese, French, you name it. It was not me being important, simply a 
    polyglot on stage.
    
    Now, when you work in our field it is impossible not to work on things related to Gell-Mann. I have however one
    important thing in common with him, the seesaw mechanism. It is natural 
    then that I speak of seesaw here. I will focus on what is the crucial question in my opinion, i.e. how to probe directly the seesaw, 
    or better to say, how to probe the origin of neutrino mass and its Majorana character. The answer
    is lepton number violation at colliders such as LHC, as Wai-Yee Keung and I suggested almost thirty years
    ago. The idea is completely analogous to a neutrinoless double beta decay: one can  produce two electrons out of 'nothing'
    if neutrinos are Majorana particles. If it was to be observed, one could use the collider determination of the scale of new 
    physics and make predictions for the neutrinoless double decay, with new physics being behind this process. Although
    this argument in favor of new physics was made about half a century ago, the myth of neutrino-less double beta decay
as a probe of neutrino mass remains to this day. I hope that this short review helps to demystify this conundrum.

\section{Introduction}

We know that neutrinos are massive but very light \cite{Strumia:2006db}. If we wish to account for tiny neutrino masses with only the Standard Model (SM) degrees of freedom, 
we need 
Weinberg's \cite{Weinberg:1979sa} $d=5$ effective operator
\begin{eqnarray}
{\mathcal L}=Y_{ij}\frac{L_iHHL_j}{M},
\label{eq:weinberg}
\end{eqnarray}
where $L_i$ stands for  left-handed leptonic doublets and  H  for the usual  Higgs doublet. This
in turn produces neutrino Majorana masses and says yes to a fundamental question raised by Majorana \cite{Majorana:1937vz}
more than seventy years ago as whether neutrinos are ``real''
particles.
 The non-renormalizable nature of the above operator signals the appearence of new physics through 
  the mass scale $M$. The main consequence is the $\Delta L=2$ violation of lepton number  through\\

a) neutrinoless double beta decay ($0 \nu 2 \beta$), suggested \cite{Racah:1937qq} soon after Majorana classic work, \\

b) same sign charged lepton pairs in colliders, suggested almost thirty years ago \cite{Keung:1983uu}.\\
 
 It is noteworthy that the conservation of lepton number was questioned already in the 30's, but it would take two
 more decades to start doubting the dogma of baryon number conservation, and turn it into an experimental question
 (for a recent discussion of this history, see \cite{Senjanovic:2009kr}).
  
  While the neutrinoless double beta decay is considered a text-book probe of Majorana neutrino mass, the like sign lepton
 pair production at colliders has only recently received
 wide attention (for a recent review and references, see for example \cite{Senjanovic:2009at}), with the arrival of the Large Hadron Collider (LHC). In what follows we will see that this process may be our best bet in probing directly the origin of neutrino 
 mass, and it can serve to make predictions for the neutrinoless double beta decay and lepton flavor violating (LFV) processes.
The point is that the neutrinoless double beta decay depends strongly on new physics accessible at LHC, and thus cannot
serve as a direct measure of neutrino mass. I cannot overemphasize this fact. The d=5 operator tells us that the standard 
model with neutrino Majorana mass is not a complete theory and thus its completion must enter in principle any physical 
effect associated with lepton number violation (LNV). Of course, if the scale M in Eq.~\eqref{eq:weinberg} is very large and thus the
new physics behind it decouples, we can safely speak of neutrino mass as the only measure of LNV, but there is no way we
can know that the scale is really large. 

It is often argued that large scales and large couplings are more natural, but that is
a wrong attitude. Large scales bring in only trouble in the form of the hierarchy issue, whereas small (Yukawa) couplings are
natural in a sense of being protected by symmetries. On top, in our world the Yukawas, at least most of them, are small and it makes
all the sense in the world to pursue this possibility, especially since it offers new physics at our reach. In the rest of this talk, I
do precisely that, and concentrate on the TeV physics, accessible to LHC, that may be behind neutrino masses.
 
  In order to get a window to that new physics, we need a renormalizable theory of the above effective operator. 
 An example is provided by the left-right ($LR$) symmetric  theory discussed in the next
section.  This is the theory that led originally to neutrino mass and the seesaw mechanism \cite{seesaw}, and as such deserves attention. Today the name seesaw usually means a completion of the SM, most of the time just by adding a set of particles which
upon being integrated out give the above d=5 operator. When one adds just one type of such states, one ends up with only
three types of seesaw, mentioned below. I will, on the other hand, focus on a theoretically motivated picture a seesaw, i.e. I
will be interested only in theories of such new states. An analogy with the generic seesaw would be an integrated 
out W boson in order to get an effective Fermi theory. Without a theory behind it, such as was provided by the SM, one gains
little, if anything through this.

 It is worth making a pause and recalling the history of the seesaw. None of the original papers simply added right-handed neutrinos
 to the standard model, first since there was no reason for it and second, since it would not help you much. This mechanism emerged
 naturally in the theories where right-handed neutrinos were a must and where neutrino mass would end up being tied to new physical
 phenomena. As such, it preceded experiment and paved the way for a true theory of neutrino mass that we all are looking for.

As we will see, if the scale of parity restoration is in the few TeV region, the theory offers a rich LHC phenomenology and a
plethora of lepton flavor violating (LFV) processes. Even more important, there is a deep connection between 
lepton number violation at LHC and in neutrinoless double decay. This important fact was recently discussed in \cite{Tello:2010am}
and here we follow it closely.
 
  The essential point is that new physics may lie behind neutrinoless double beta decay.  This suggestion was made more than fifty years ago 
  \cite{feinbergbrown}, and some thirty years ago it was argued that this may happen in left-right symmetric theories \cite{Mohapatra:1980yp}. And yet, it is 
 so often claimed that this process is a probe of neutrino mass, that it is crucial to give a clear example of a theory that may say the opposite. 
 This is the central aspect of \cite{Tello:2010am}, where it is shown how LR
 symmetry at the LHC scale is likely to dominate over neutrino mass as the source of neutrinoless double beta decay.  Needless to say, this
 is not the only new physics that could be behind neutrinoless double beta decay. Another logical possibility is the minimal supersymmetric 
 standard model \cite{Allanach:2009iv}. Due to a large number of parameters of the MSSM, it is much harder though to make 
 predictions in this theory. 
 
  It is also possible that neutrino mass lies behind the neutrinoless double decay, with the new physics responsible for neutrino mass
   accessible at LHC. This was studied  \cite{Kadastik:2007yd} in the context of the type II seesaw \cite{typeII}, and also in the case of a simple SU(5) grand unified 
   theory \cite{Arhrib:2009mz} which predicts  \cite{Bajc:2006ia} a hybrid type I \cite{seesaw} plus type III seesaw \cite{Foot:1988aq}. 
   
   The SU(5) theory is 
   particularly appealing since it predicts a light fermion triplet, responsible for the type III seesaw, with a mass below TeV.
   It is through its decays that one can reconstruct the neutrino mass matrix, for it turns out that the lightest neutrino is
   effectively massless. Thus you have seesaw mechanism predicted by an underlying theory, and not just put by hand,
    and furthermore the theory itself fixes the seesaw scale to be at the LHC energies. This
   should serve as a prototype of a theory of the neutrino mass origin one is after. Since it has been reviewed in \cite{Senjanovic:2009at}, we rather focus on the left-right symmetric theory. Although it does not predict its own scale,  the necessary presence
   of right-handed neutrinos and the connection of neutrino mass to the scale of parity restoration make this theory special. It
   forced its way to massive neutrinos when most people believed in the massless one, suggested by the standard model. In the Summary and Outlook we discuss
   though some essential features of the SU(5) theory and its prospects for the LHC, for the sake of comparison with the LR symmetric theory. 
  
    One possible direction of getting a handle on the LR symmetry is a SO(10) grand unified theory, where this symmetry
    is gauged in a form of charge conjugation. It is a highly suggestive and  appealing theory, but I do not discuss it here
    for it predicts the LR scale to be enormous, far above the LHC reach which is the topic of my talk. For a review and
    references of this approach, see \cite{lr scale}.
  
 \section{Left-right symmetry and the origin of neutrino mass}

 The idea of LR symmetry comes as a desire to understand the origin of parity violation in weak
interactions. It is important to recall that a wish to have parity as a fundamental symmetry in beta
decay is as old as the suggestion of its breakdown. In their classic paper, Lee and Yang~\cite{Lee:1956qn} argue in
favor of the existence of the opposite chirality heavy proton and neutron, which would make the
world parity symmetric at high energies. 

\paragraph{Mirror fermions.}In the modern SM language, these are coined mirror fermions
and there are a number of important theoretical frameworks that imply them: Kaluza-Klein theories~\cite{Witten:1981me}, family unification based on large orthogonal groups~\cite{GellMann:1980vs,Senjanovic:1984rw,Bagger:1984rk}, N=2 supersymmetry~\cite{delAguila:1984qs}, some unified models of gravity~\cite{GGUT}. Mirror fermions appear naturally in the simplest and most physical way of gauging baryon and lepton number symmetry.

 I take the pain of discussing the mirrors, a topic outside of the scope of this review, in order to emphasize yet another topic that Murray Gell-Mann worked on~\cite{GellMann:1980vs}.  One would imagine that with the high precision data their existence would be ruled out.
 Surprisingly, one can still have a mirror family per each ordinary one~\cite{Martinez:2011ua}, as long as as there is another scalar doublet. Moreover, the second doublet is forced to be quite inert~\cite{Deshpande:1977rw} and light, thus
 becoming a possible dark matter candidate~\cite{Barbieri:2006dq, DM}. The usual Higgs boson must be quite heavy,
 on the order of 450-500 GeV, fitting perfectly with strongly boosted gluon fusion production. The narrow parameter
 space of this appealing possibility makes it exciting, and LHC will reveal soon whether it is true or not.

\paragraph{Left-right gauge theory.}The LR symmetric gauge theories, on the other hand, keep the fermionic content of the SM intact, and instead double the weak gauge sector. The minimal such theory ~\cite{leftright} is based on the following gauge group
(suppressing colour):
$$
G_{LR}= SU(2)_L \times SU(2)_R \times U(1)_{B-L}\,,
$$
plus a symmetry between the left and right sectors.  Quarks and
leptons are completely LR symmetric
\be
Q_{L,R }= \left( \begin{array}{c} u \\ d \end{array}\right)_{L,R}\,,\qquad
\ell_{L,R} = \left( \begin{array}{c} \nu \\ e \end{array}\right)_{L,R}.
\label{ds21}
\ee

The formula for the electromagnetic charge becomes
\begin{equation}
Q_{em} = I_{3 L} + I_{3 R} + {B - L \over 2}\,.
\label{ds22}
\end{equation}

The Higgs sector consists of the following multiplets
\cite{triplet}: the bi-doublet $\Phi \in (2_L,2_R,0)$ and the
$SU(2)_{L,R}$ triplets $\Delta_L \in (3_L,1_R,2)$ and $\Delta_R \in
(1_L,3_R,2)$, according to the $SU(2)_L \times SU(2)_R \times
U(1)_{B-L}$ quantum numbers
\begin{equation}
\Phi = \left[\begin{array}{cc}\phi_1^0&\phi_2^+\\\phi_1^-&\phi_2^0\end{array}\right]
\;
\Delta_{L, R} = \left[ \begin{array}{cc} \Delta^+ /\sqrt{2}& \Delta^{++} \\
\Delta^0 & -\Delta^{+}/\sqrt{2} \end{array} \right]_{L,R}
\label{ds32}
\end{equation}

It can be shown that the first stage of the breaking of the $G_{LR}$
down to the SM model symmetry, takes the following
form~\cite{Mohapatra:1980yp}
\begin{equation}
\langle \Delta_L\rangle = 0 \;\;\; , \;\;\; \langle \Delta_R\rangle = \left[ \begin{array}{cc} 0 & 0 \\
v_R & 0 \end{array}\right]
\label{ds34}
\end{equation}
At the next stage, the neutral components $\Phi$ develop a VEV and break the SM symmetry down to $U(1)_{em}$
\begin{equation}
 \langle \Phi\rangle = \left[ \begin{array}{cc} v_1 & 0 \\
0 & v_2\, {\rm e}^{i\a}\end{array}\right]
\label{ds35}
\end{equation}
where $v_{1,2}$ are real and positive, $M_W^2 = g^2 v^2\equiv g^2
(v_1^2 + v_2^2)$ and $g\equiv g_L=g_R$ denote the SU(2) gauge
couplings. In turn, $\Delta_L$ develops a tiny VEV 
$\langle\Delta_L\rangle\propto v^2/v_R$. This will be crucial for the light neutrino masses
(see below).

\paragraph{Gauge bosons.} The gauge boson masses are given by
\begin{eqnarray}
& M_{W_R}^2 \simeq   g^2 \, v_R^2  \\
& M_{Z_R}^2 \simeq    2 ( g^2 + g_{B-L}^2 ) \,
v_R^2=\frac{2g^2}{g^2-g_Y^2}M_{W_R}^2\simeq 3M_{W_R}^2\,,
\end{eqnarray}
where we used the relation $g_Y^{-2}=g^{-2}+g_{B-L}^{-2}$ among $g_Y$
and $g_{B-L}$, respectively the gauge couplings of $Y/2$ and
$(B-L)/2$. In other words, $M_{Z_R}\simeq 1.7 M_{W_R}$ and the limit
on $W_R$ becomes especially important if one wishes to discover also
$Z_R$ at LHC.

In the above, we neglected the tiny mixing among left and right gauge bosons.
 Although in
general these mixings could play an important role, for a large LR
scale they obviously become secondary or irrelevant. 
For a recent detailed discussion of the limits on the LR scale, spectrum of the theory and the associated phenomenology,
  see~\cite {Maiezza:2010ic}, \cite{Zhang:2007da}, \cite{Beall:1981ze}. The bottom line is a theoretical lower limit $M_{W_R} \gtrsim 2.5 \mbox{ TeV}$ 
in the minimal model to which we stick. Interestingly enough, the early LHC data can already 
be used to set a direct search limit $M_{W_R} \gtrsim 1.4 \text{ TeV}$ in a big portion of
the parameter space of right-handed neutrino masses~\cite{Nemevsek:2011hz}. Even in the
opposite case of Dirac neutrino masses (or light right-handed neutrinos), when the
$W_R \to \ell + \slashed{E}$, one obtains roughly the same limit~\cite{WRmissingE}. Thus,
experiment is finally closing up on the theoretical bound, and soon we can anticipate a discovery?
 
 When needed we will choose a representative point 
$M_{W_R}=3.5 \mbox{ TeV}$, which makes the LR symmetry accessible to LHC (see below).

The symmetric Yukawa couplings of the triplet relevant for our discussion are
\begin{equation}
	{\cal L}_Y (\Delta) = \frac{1}{2} \ell_L  \frac{M_{\nu_L}}{\langle \Delta_L\rangle}
	 \Delta_L \ell_L +
	\frac{1}{2} \ell_R  \frac{M_{\nu_R}}{\langle \Delta_R\rangle}  \Delta_R \ell_R+ 
\text{h.c.}\,,
\label{eqLDelta}
\end{equation}
where $M_{\nu_L}$ and $M_{\nu_R}$ are Majorana mass matrices of light and heavy neutrinos. In principle there are also
Dirac Yukawa couplings connecting the two. When these tiny couplings play a negligible role, the resulting seesaw is
called type II~\cite{typeII}. For the sake of illustration, and without loss of generality, in what follows we stick to this
appealing scenario. This is done in order to demonstrate the connection between the LHC and the low energy experiments,
such as $0 \nu 2 \beta$ and LFV. The point is that LHC can measure the right-handed leptonic mixing matrix $V_R$, which is
needed in order to make predictions for the latter processes. In the absence of this, meanwhile, we will show how the 
knowledge $V_R$ suffices, together with the new particle masses, to make clear statements about low energy experiments.
Since we do not have the information of $V_R$ as of yet, we will take a version of the minimal theory that through the LR symmetry relates the left and right mixing angles. As we discuss below, one ends up with a 
prediction $V_R = V_L^*$.  

 The bottom line is, once the right-handed sector is integrated
 out, the fact that light neutrinos are Majorana particles. The smallness of neutrino mass is the consequence
 of near maximality of parity violation in beta decay, and in the infinite limit for the $W_R$ mass
 one recovers massless neutrinos of the SM. This is what we are after: a theory where neutrino mass
 is related to new physics. I cannot over stress the fact that that this theory proceeded experiment: neutrino
 mass and the seesaw mechanism were suggested in LR theories long before neutrino oscillations 
 established non-vanishing neutrino masses.

\paragraph{LR symmetries.} The pattern of mass matrices depends on the
kind of Left-Right symmetry imposed on the model in the high-energy,
symmetric phase. It is easy to verify that the only realistic discrete
symmetries exchanging the left and right sectors, preserving the kinetic terms  are
%
\be \P:\left\{\ba 
Q_L\leftrightarrow Q_R\\
[.7ex]\Phi\leftrightarrow \Phi^\dag\\
[.7ex]\Delta_L \leftrightarrow \Delta_R
\ea\right.
\qquad
\C:\left\{\ba  Q_L\leftrightarrow (Q_R)^c\\
[.7ex]\Phi \leftrightarrow \Phi^T\\
[.7ex]\Delta_L \leftrightarrow \Delta_R^*
\ea\right.
\label{eq:PCdef}
\ee
where $(Q_R)^c=C\g_0Q_R^*$ is the charge-conjugate spinor.

The names of \P\ and \C\ are motivated by the fact that they are
directly related to parity and charge conjugation, supplemented by the
exchange of the left and right SU(2) gauge groups, as is evident from
(\ref{eq:PCdef}).

Note that $(Q_R)^c$ is a spinor of left chirality like $Q_L$, and thus
\C\ has an important advantage: since it involves the spinors with
same final chirality, it can be gauged, i.e. it allows to have this
symmetry embedded in a local gauge symmetry. In fact, in the SO(10)
grand unified theory \C\ is a finite gauge transformation. The gauging
 not only provides an aesthetic advantage, it guarantees the protection from
unknown high energy physics, gravitational effects, etc.

In spite of this, the simpler case of \P\ was the main subject of past
investigations~\cite{leftright}, probably for historical reasons, since
the original papers used it.  The case of \C\ on the other hand was
not extensively studied, at least not in the context of
phenomenology. For the reasons discussed above, we opt here for \C\
as the LR symmetry in what follows, but this brings no less of generality
on what follows. Similar relations would be obtained if one were to use 
\P\, and the reader should understand the choice of \C\ only as an example.
As shown in \cite{Maiezza:2010ic}, in this case the true phenomenological
limit on the LR scale in the minimal theory is
$M_{W_R}\gtrsim2.5\,\TeV$.  
 
In order to make phenomenological predictions in this theory, we need to know $V_R$ as we stressed 
repeatedly. This will hopefully be provided by the LHC, once $W_R$ is discovered. Meanwhile, in order
to exemplify the power of the knowledge of $V_R$ we take a possibility of type II seesaw.
Due to $\mathcal C$, the theory is then characterized by the
proportionality of the two neutrino mass matrices
%
$M_{\nu_R}/\langle \Delta_R \rangle = M_{\nu_L}^*/\langle \Delta_L \rangle^*$.
%
An immediate important consequence is that the mass spectra are proportional to each other
\begin{equation}
	m_N \propto m_\nu\,,
\label{spectrum}
\end{equation}
where $m_N$ stands for the masses of the three heavy right-handed neutrinos $N_i$ and $m_\nu$ for those of the
three light left-handed neutrinos $\nu_i$.  In this theory, there are both left and right-handed charged gauge
bosons with their corresponding leptonic interactions in the mass eigenstate basis:
\begin{equation}
\mathcal L_{W} = \frac{g}{\sqrt{2}} \left(
\bar \nu_L V_{L}^\dag  \slashed{W}\!_L e_L +
\bar N_R  V_{R}^\dag \slashed{W}\!_R e_R\right)
+\text{h.c.}\,.
\end{equation}
Since the charged fermion mass matrices are symmetric (due to the symmetry under $\mathcal C$), one readily
obtains a connection (up to complex phases, irrelevant to our discussion) between the right-handed and the
left-handed (PMNS) leptonic mixings matrices
\begin{equation}
	V_R = V_L^*.
	\label{rightmixing}
\end{equation}
  
We wish to pause here and make sure that our message is carried through. The above relation is valid only 
for the case of the type II seesaw, and it cannot be taken as a prediction of the theory. It should be viewed as
an example of what LHC can achieve for us if $W_R$ is found and one is able to measure $V_R$, for the rest will
follow as described below. It is essential that (\ref{rightmixing}) can be probed and it can be taken as a test of
the type II mechanism.

Before turning to a discussion of the lepton flavor violation in this theory, a comment is called for 
regarding a notorious domain wall problem, a result of the spontaneous symmetry breaking of
discrete symmetries, such as $\mathcal C$ or $\mathcal P$. A possible way out~\cite{Dvali:1995cc}
 is a non-restoration of symmetry at high 
temperature \cite{Weinberg:1974hy}, or a tiny breaking of these symmetries by say Planck scale 
suppressed effects~\cite{Rai:1992xw}.
 
\paragraph{Lepton Flavor Violation.} Lepton flavor violation in LR symmetric theories has been the subject of interest from the 
beginning of the model based on the seesaw mechanism ~\cite{peter},~\cite{Mohapatra:1980yp} and was studied in detail    
in~\cite{Cirigliano:2004mv}. What is new in \cite{Tello:2010am} is the connection with LHC and especially the
quantitative implications for \BB. I give here the limits on the masses of right-handed neutrinos, relevant for
predictions regarding the neutrinoless double beta decay. Needless to say, if the LHC were to measure their
masses and mixings, one could in turn make predictions for LFV processes. 
 
\begin{figure}
	\includegraphics[width= 4cm]{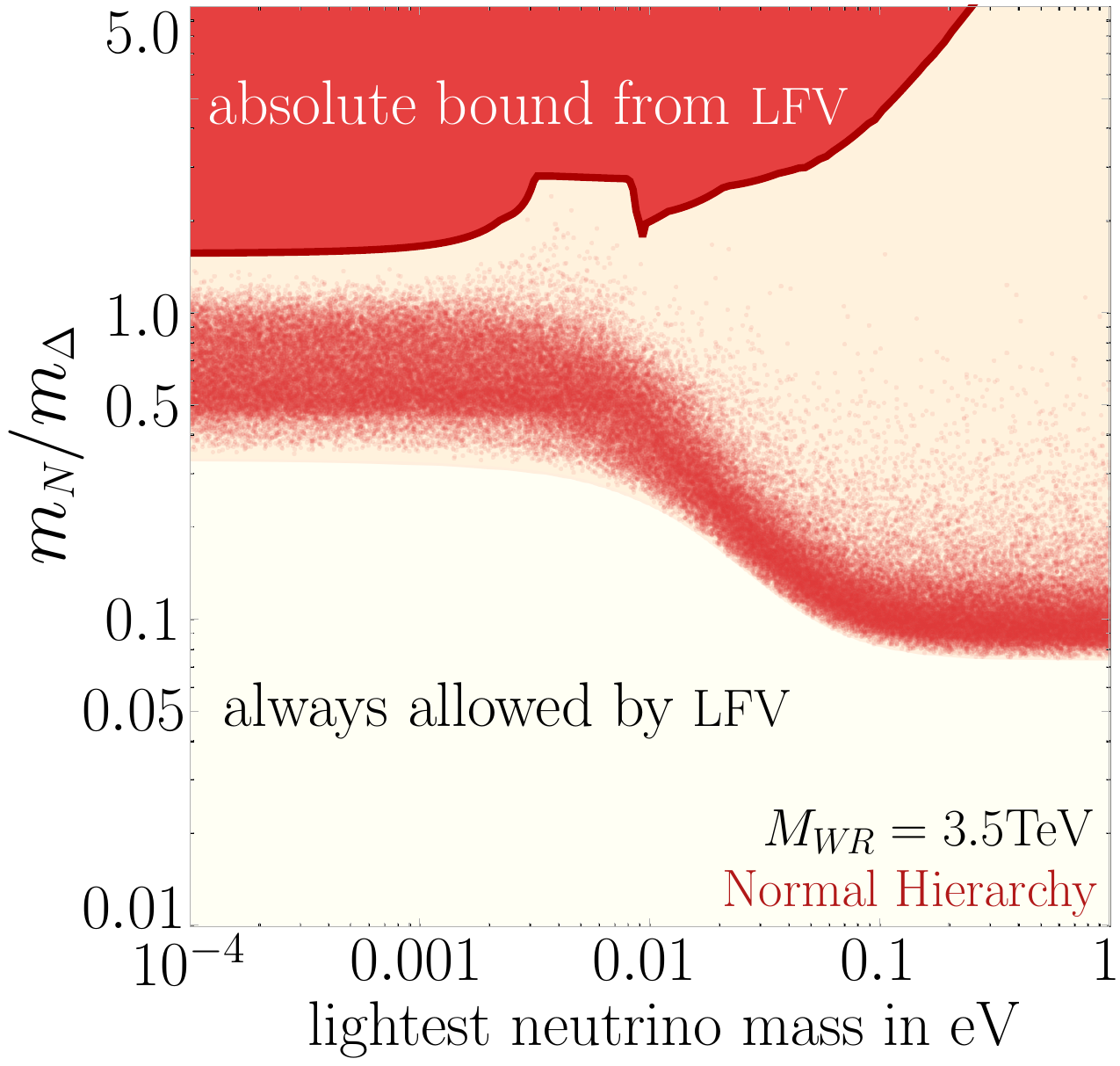}
	\includegraphics[width= 4cm]{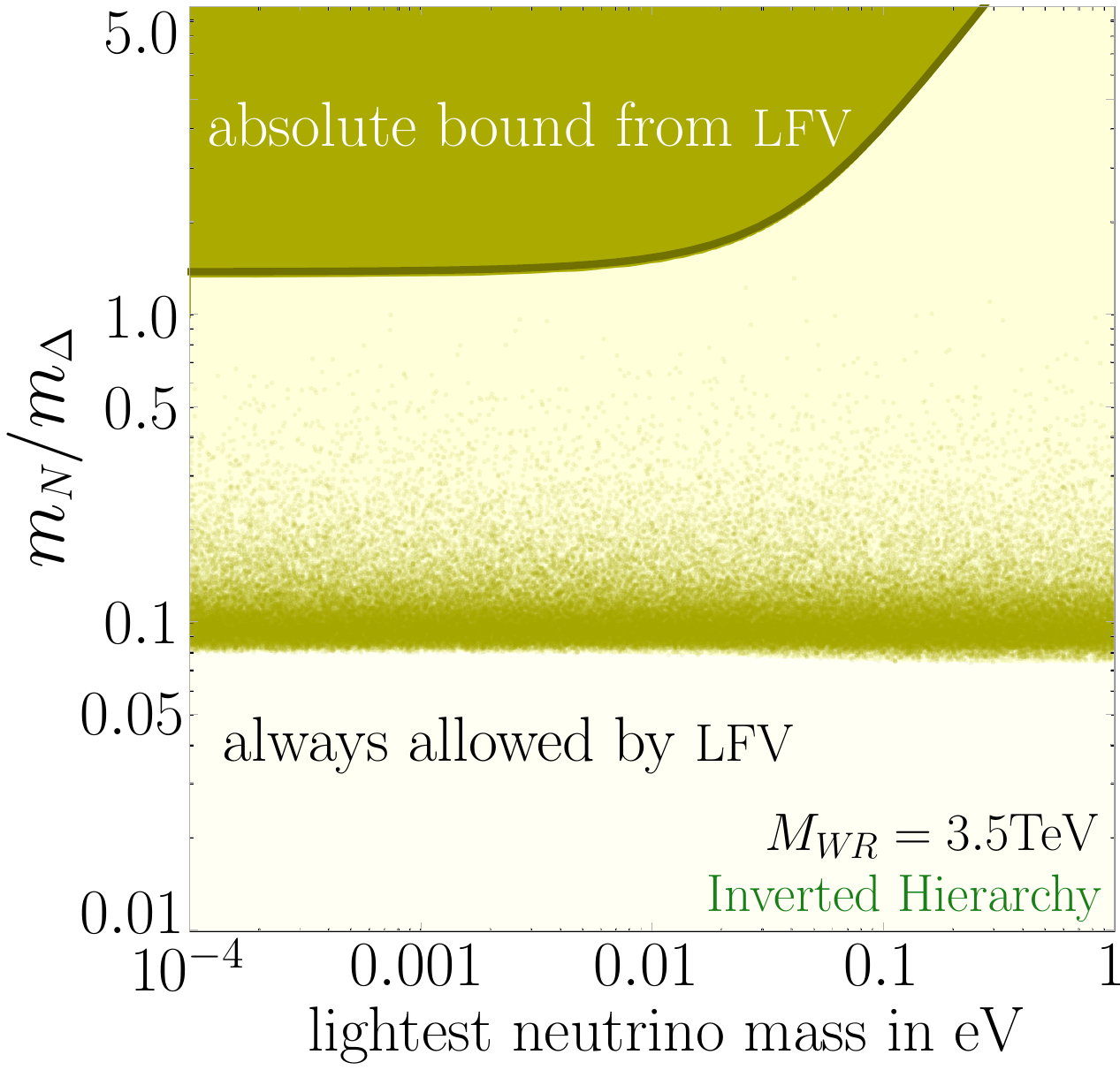}%
        \vspace*{-1ex}%
	\caption{\it Combined bounds on $m_N^\text{heaviest}/m_{\Delta}$ from LFV, taken from \cite{Tello:2010am}.  The dots show the (most
          probable) upper bounds resulting for different mixing angles, Dirac and Majorana phases (varied
          respectively in the intervals $\{\theta_{12}, \theta_ {23}, \theta_{13}\}=$ $\{31 \text{-}39^{\circ}
          ,37 \text{-}53^{\circ}, 0 \text{-}13^{\circ}\}$ and $\{0,2\pi\}$).  The dark line is the absolute
          upper bound. The plot scales as $M_{W\hspace{-0.05cm}R}/3.5\,\TeV$.}
        \vspace{-0.1cm}%
        \label{figLFV} 
\end{figure}

There are various LFV processes providing constraints on the masses of right-handed neutrinos and doubly
charged scalars illustrated in Fig.~\ref{figLFV}.  It turns out that $\mu \rightarrow 3 e$, induced by the
doubly charged bosons $\Delta_{L}^{++}$ and $\Delta_{R}^{++}$, provides the most relevant constraint and so we
give the corresponding branching ratio (see \cite{Tello:2010am})
\begin{equation}
\!\!\!\!{\text{BR}}_{\mu \rightarrow 3 e} = \frac{1}{2}\!  \left(\!\frac{M_W}{M_{W_R}}\!  \right)^{\!\!4} \left|V_L \frac{m_N}
{m_{\Delta}}V_L^T \right|^2_{e \mu} \left|V_L \frac{m_N}{m_{\Delta }}V_L^T \right|^2_{ee},
\end{equation}
where $1/m_{\Delta}^2\equiv1/m_{\Delta_L}^2+1/m_{\Delta_R}^2$. The current experimental limit is $\text{ 
BR}(\mu\rightarrow 3e)<1.0 \times 10^{-12}$~\cite{Bellgardt:1987du}. 

The LFV transition rates become negligible when the masses of $M_{W_R}$ and $m_{\Delta}$ become
larger than about 100\,\TeV.  We are interested in LHC accessible energies,  
  in which case the smallness of the LFV is governed by the ratio $m_N/m_\Delta$, in addition to
mixing angles. In this sense LFV is rather different from LNV which in oder to be observable needs roughly
a TeV scale. It is perfectly possible that the LR scale, much above the LHC reach, leads to observable LFV
processes; however, it would be basically impossible to verify that. This is why the LHC scale new physics 
becomes so important, for it would relate all these different processes.  The reason that it is still possible
not to be in conflict with the LFV experimental limits, even with the TeV scale LR symmetry, is of course 
the fact that the mixings and phases, together the masses of N's can control the size of the relevant rates.
The crucial dimensionless parameter is $m_N^{heaviest}/m_{\Delta}$, and in \cite{Tello:2010am} we have
plotted the upper bound on this quantity, varying the mixing
angles and phases (LFV plots also take into account $\mu\rightarrow e$ conversion in Au
  nuclei~\cite{Bertl:2006up}, $\mu \rightarrow e \gamma$~\cite{Brooks:1999pu} and rare $\tau$ decays
  such as $\tau\rightarrow 3\mu$, etc.~\cite {tauLFV}) (see fig.~\ref{figLFV}, taken from \cite{Tello:2010am}).
  An immediate rough consequence seems to
follow: $m_N^{\text{heaviest}} / m_{\Delta} < 0.1$ in most of the parameter space. However, the strong dependence on
angles and phases allows this mass ratio up to about one in the case of hierarchical neutrino
spectra. This serves as an additional test at colliders of type II seesaw used here. For
degenerate neutrinos, unfortunately, no strict constraint arises: see again Fig.~\ref{figLFV}.

It is worth comparing the LFV in his theory with the SU(5) one \cite{Bajc:2006ia} discussed in the introduction. Whereas 
here LFV is expected to be large and the theory needs to shield itself from its excess, in the latter case you expect at first glance
negligible amount of LFV. Namely, as we remarked in SU(5) neutrino mass stems from a light fermion triplet (type III seesaw),
and for generic values of the triplet Yukawa couplings these rates are much below experimental sensitivities. However, the situation
may be more subtle; a careful, detailed study cab be found in~\cite{Kamenik:2009cb}. Observing LFV will thus not shed light by
itself on the theory behind, but can only serve as a complementary check of a theory in question.
 
Before closing this section, we wish to remark on an exciting possibility of planned new experiments \cite{Ankenbrandt:2006zu},
\cite{jpark} on 
$\mu \to e$ conversion, that could improve the sensitivity by four to six orders in magnitude. If a signal is observed,
one can in principle measure the CP violation phases of $V_R$ \cite{Bajc:2009ft} that enter into
the other LFV processes, and especially into the neutrinoless double beta decay.  This could serve as a 
check of the theory and the role of LFV would change 
drastically, for one could start probing the theory behind the LFV.

  \section {Neutrinoless double beta decay}  
 
\begin{figure}
\begin{center}
\includegraphics[scale=0.8]{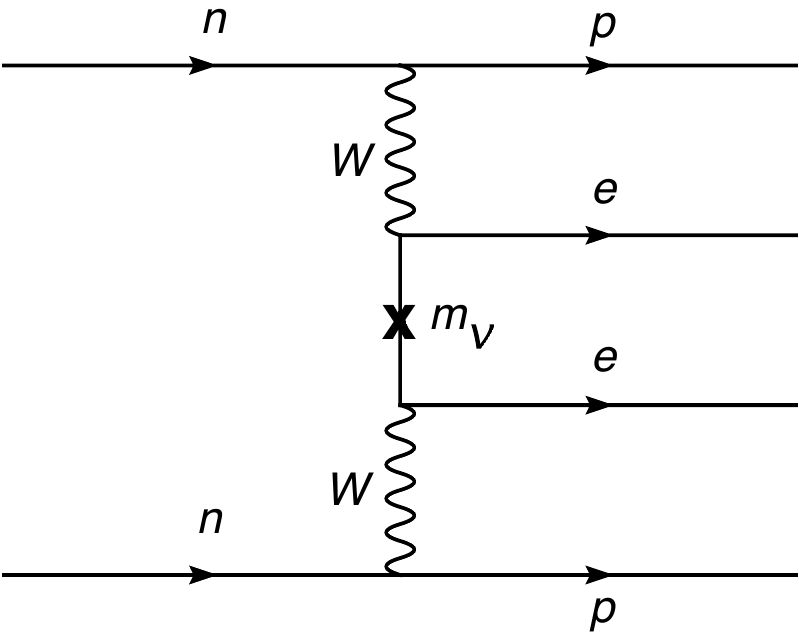} 
\caption{Neutrino-less double beta decay through the neutrino Majorana mass.}\label{bb0v}
\end{center}
\end{figure}

%
%
As discussed in the Introduction, although often claimed, 
  in general neutrino Majorana mass is not directly connected to neutrinoless double beta decay. While it does produce it (see 
 Fig.~\ref {bb0v} ), the inverse is 
not true. Neutrinoless double beta decay does not imply the measure of neutrino mass, since it depends on the completion of
the SM needed for the above $d=5$  operator in (\ref{eq:weinberg}). 

\begin{figure}[h!]
\begin{center}
\includegraphics[scale=0.8]{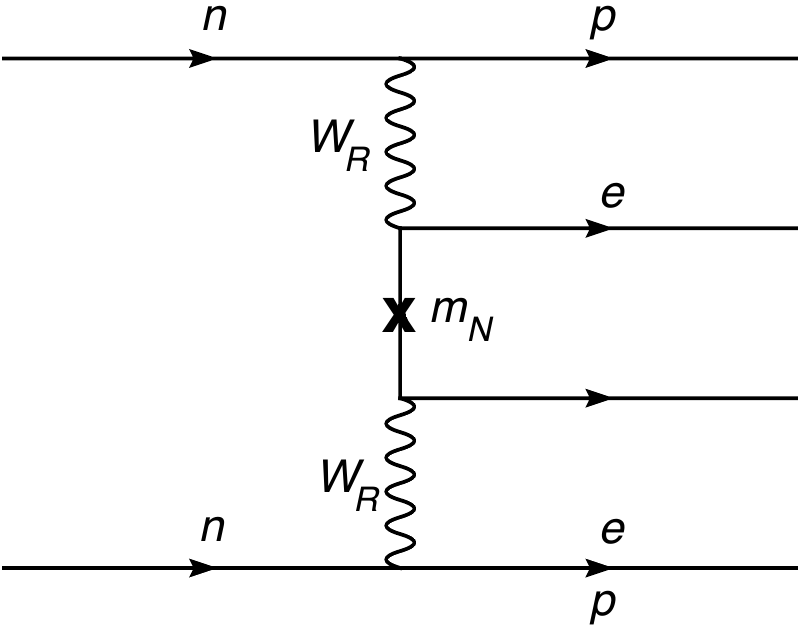} 
\caption{Neutrino-less double beta decay induced by the right-handed gauge boson and right-handed neutrino.}
\label{bb0v2}
\end{center}
\end{figure}

The LR theory gives a new contribution to the neutrinoless double beta decay, through the 
right-handed sector, as in Fig.~\ref {bb0v2}. This was discussed originally a long time ago ~\cite{Mohapatra:1980yp},
and was used as an argument for boosting a search for this process.  

It gives for the $0 \nu 2 \beta$ transition amplitude 
\begin{equation}
\mathcal A_{\text{RR}}\propto \frac{1}{M_{W_R}^4}\left(  \frac{1}{m_N} \right)^{ee}
\end{equation}
to be compared with the usual $W$ contribution
\begin{equation}
\mathcal A_{\text{LL}} \propto \frac{1}{M_{W}^4} \frac{m_\nu^{ee}}{p^2} 
\end{equation}
where $m_\nu^{ee}$ is the 1-1 element of the neutrino mass matrix $m_\nu$ and $p \approx 100\,\MeV$
a measure of the neutrino virtuality.

We have
\begin{equation}
\frac{\mathcal A_{\text{RR}}}{\mathcal A_{\text{LL}}} \simeq \left( \frac{M_{W}}{M_{W_R}}\right)^4 \frac{p^2}{m_\nu^{ee}}\left(  \frac{1}{m_N} \right)^{ee}
\end{equation}
 
 With $W_R$ mass in the TeV region and the right-handed neutrino masses ($m_N = m_{\nu_R}$) in the 100 GeV region, 
 this contribution can easily dominate over
the left-handed one. Light neutrino mass can even go to zero while keeping the $W_R$ contribution finite. 
  
\begin{figure}
	\includegraphics[width=4cm]{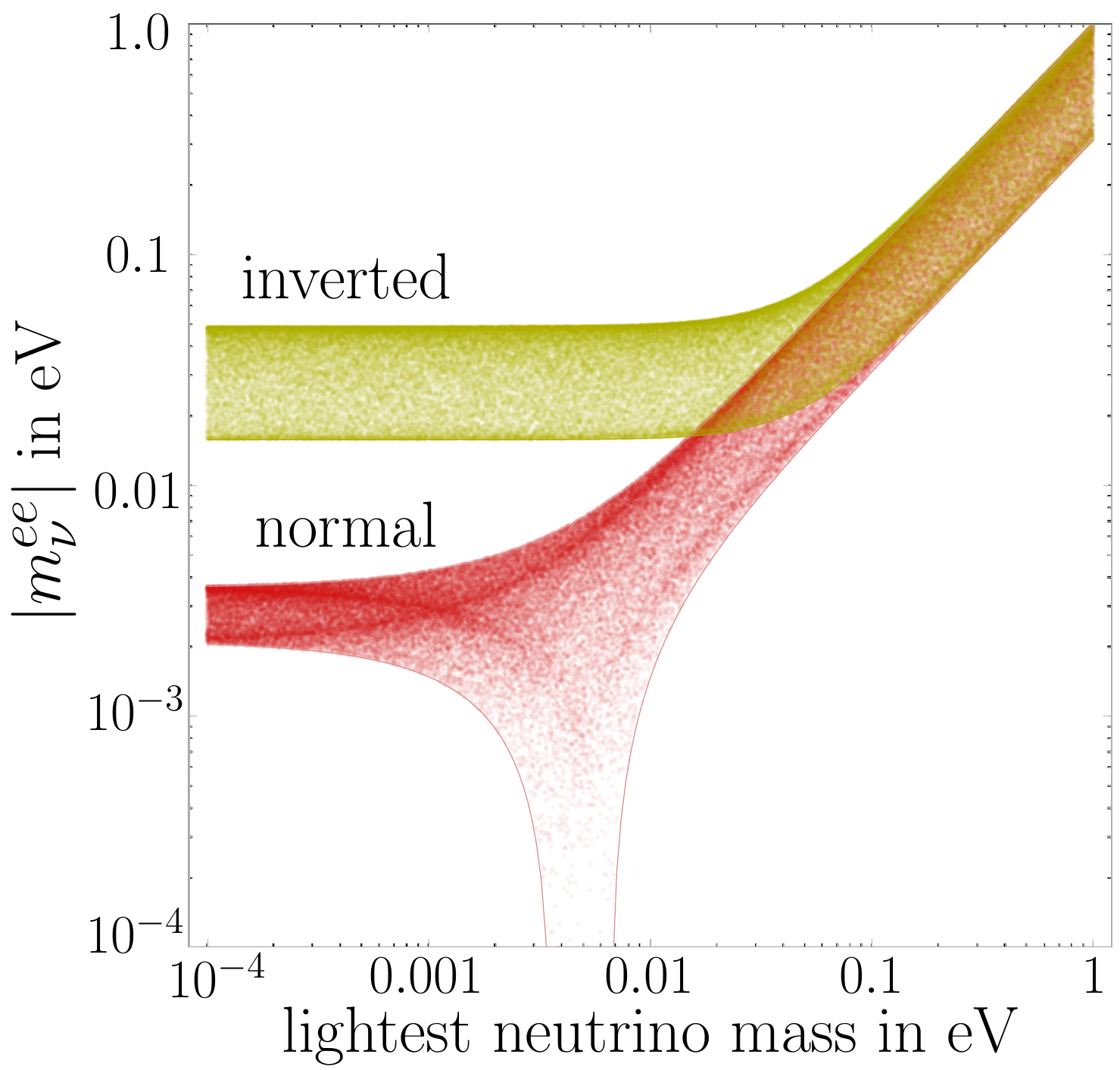}
	\includegraphics[width=4cm]{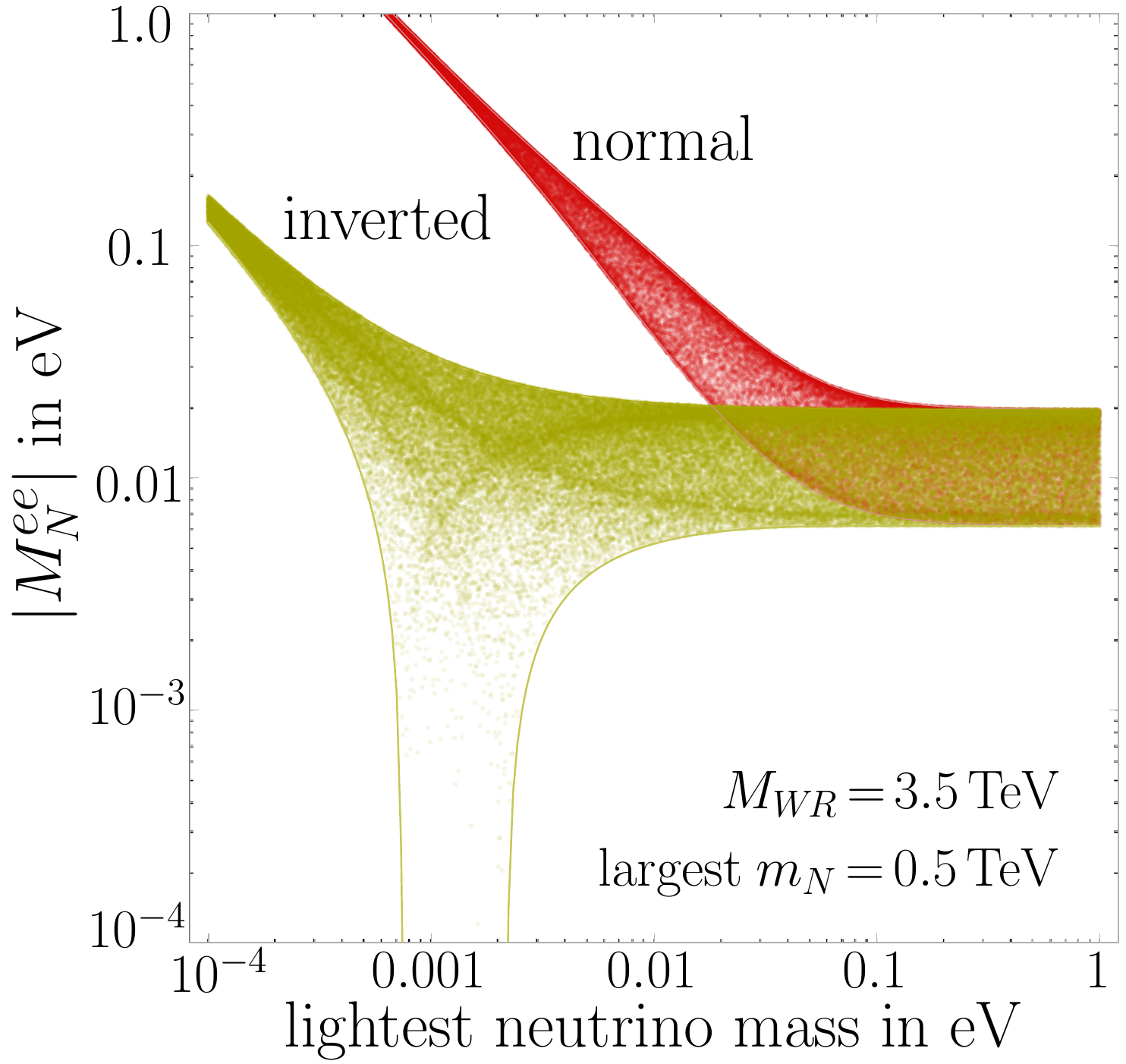}%
        \vspace*{-1ex}%
	\caption{\it Neutrinoless double beta decay. The canonical contribution (left) from light neutrino mass and the new physics
          part (right), with $|M_{\scriptscriptstyle {N} }^{ee}|$ defined in Eq.~\eqref{eqMNee}.
          The mixing angles are fixed at $\{\theta_{12}, \theta_{23}, \theta_{13}\} = \{35^{\circ}
          ,45^{\circ},7^{\circ}\}$, while the Dirac and Majorana phases vary in the interval
          $\{0,2\pi\}$.  This figure is taken from \cite{Tello:2010am}. }
        \vspace{-2ex}
        \label{figTelloPlot}%
\end{figure}

  In other words, $W_R$ at LHC suggests strongly that new physics may dominate $0 \nu 2 \beta$ as argued originally
  in  \cite{Mohapatra:1980yp}. What is remarkable is that the opposite is true, too: the new physics as a source of $0 \nu 2 \beta$
should be accessible to LHC in order to do the job. This is evident if one writes the new physics contribution in a natural form
%
\begin{equation}
	\mathcal A_{\text{NP}} \propto G_F^2 \frac{M_W^4}{\Lambda^5},
\end{equation}
where $\Lambda$ is the scale of new physics.  Compare this with the conventional neutrino mass source of $0 \nu 2 \beta$,
which we rewrite slightly 
\begin{equation}
	\mathcal A_\nu \propto G_F^2 \frac{m_\nu^{ee}}{p^2}.
\end{equation}
 Clearly, the new physics enters the game at $\Lambda \sim\TeV$.
  This fact alone provides a strong
motivation to pursue this line of thought.
 
This is quite different from a case when neutrino mass lies behind 
$0 \nu 2 \beta$,  and these two programs are not to be confused.

  In what follows we neglect the tiny $W_L$-$W_R$ mixing of
$\mathcal{O}(M_W/M_{W_R})^2 \lesssim 10^{-3}$ and contributions coming from the bidoublet through
the charged Higgs, because of its heavy mass of at least 10\,\TeV~\cite{Maiezza:2010ic}. In this case
we are left with only two extra contributions and with an effective Hamiltonian given by (the
contribution from the left-handed triplet is completely negligible)
\begin{equation} \label{eq:0nubblag}
\!\!\!\mathcal{H}_{\text{NP}}= G_F^2 V_{Lej}^2 \! \left[ \frac{1}{m_{N_j}}\!  +\! 
\frac{2 \ m_{N_j} }{ m^{2}_{\Delta_R^{++}} }  \right]\! \frac{M_W^4}
{M_{W_R}^4} \! J_{R\mu }^{\,}J^{\mu}_R\, \overline{e_{R}} e^{\,\,c}_R \,,
\end{equation}
where $J_{R \mu}$ is the right-handed hadronic current.
Making use of the LFV constraint  $ m_N/m_{\Delta} \ll 1$ one can neglect the $\Delta_R^{++}$ contribution 
  and write the total decay rate as
  \begin{equation}
\frac{\Gamma_{  0\nu\beta\beta }}{\text{ln\,2}} = G \cdot \left| \frac{\mathcal{M}_{\nu}}{m_e}\right|^2   \Bigg( |  
m_{\nu}^{ee} |^2 + \Bigg| p^2 \frac{M_W^4}{M_{W_R}^4}  \frac{V_{Lej}^2}{m_{N_j}}   \Bigg|^2 \Bigg)\,,
\end{equation}
where $G$ is a phase space factor, $\mathcal{M}_{\nu}$ is the nuclear matrix element relevant for
the light neutrino exchange, while $p$ measures the neutrino virtuality and accounts also for the
ratio of matrix elements of heavy and light neutrinos.

In order to illustrate the impact of the Dirac and Majorana phases on the total decay rate, we show in the
left frame of Fig.~\ref{figTelloPlot} (taken from \cite{Tello:2010am}) the well known absolute value of $m^{ee}_{\nu}$ 
which measures the standard neutrino mass contribution \cite{Vissani:1999tu}, while the
corresponding effective right-handed counterpart,
\begin{equation} \label{eqMNee}
	M_N^{ee} = p^2 (M_W/M_{W_R})^4 V_{Lej}^2/m_{N_j}\,,
\end{equation}
is shown separately in the right frame.  This plot has been made using Eqs.~\eqref{spectrum} and \eqref{rightmixing}
with $p = 190\,\MeV$ and taking the entire range of $V_L$ to be allowed by LFV, see
Fig.~\ref{figLFV}.

A striking feature which emerges is the reversed role of neutrino mass hierarchies. While in the case of
neutrino mass behind neutrinoless double beta decay the normal hierarchy matters less and degeneracy is most
promising, in the case of new physics it is normal hierarchy that dominates and degeneracy matters less. Even 
more striking is a situation in the far left corner, when the mass of the lightest neutrino species becomes smaller
and smaller. This region is interesting for cosmological considerations which keep lowering the sum of neutrino masses. Moreover, recent studies of the BBN seem to be pointing towards 
four (even five) light neutrino species \cite{Hamann:2010bk} with masses in the sub-eV region. Four light neutrino species at the BBN
would force the lightest right-handed neutrino to lie in the sub-eV region, which, from  (\ref{spectrum}), would imply effectively 
massless lightest neutrino. Notice that in this theory the light-right handed neutrino is almost as equally abundant as the left-handed species, for it decouples very late (in the case of sterile neutrinos, without gauge interactions, one has to rely on tiny Yukawa couplings, a long shot. 

 In the case of the standard neutrino mass source of the $0 \nu 2 \beta$, this portion
of the parameter space is hopeless in the case of normal hierarchy, with some hope for the inverse hierarchy,
if the experiments get bellow 0.1 eV for $m^{ee}_{\nu}$. On the contrary, with the new physics of $W_R$ being
the culprit, the situation is highly favorable, and the present experimental situation already sets strong limits
on the masses of the other two right-handed neutrinos. This can be great news for this theory, and could serve
as a crucial check of its validity.  

\begin{figure}[t]
	\includegraphics[height=.65\columnwidth]{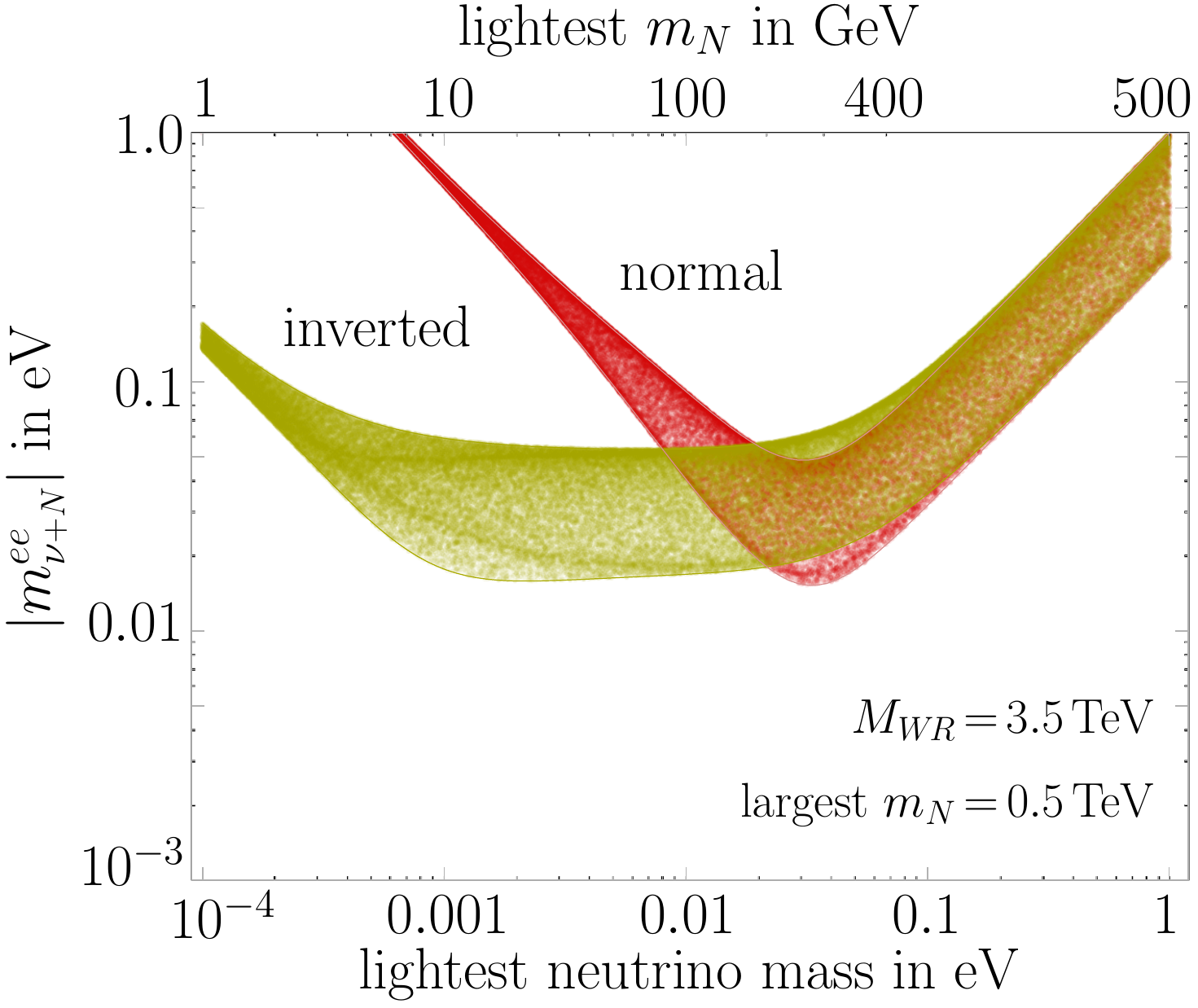}\hspace*{1em}%
       \vspace*{-1ex}%
	\caption{\label{figbetaLR}\it
	Effective \BB\ mass parameter $|m_ {\scriptscriptstyle { \nu +N } }^{ee}|$, 
	a measure of the total \BB\ rate including contributions from both left and right currents. This figure is given in \cite{Tello:2010am}.}   
     \vspace{-0.1cm}
\end{figure}

The total \BB\ rate is governed by the effective mass parameter
\begin{equation}
|m_ {\nu + N}^{ee}|= (| m_\nu^{ee}|^2 + | M_N^ {ee}|^2)^{1/2}
\end{equation}
i.e.\ a quantity that supersedes the standard matrix element $m^{ee}_{\nu}$ in the parameter space
accessible to LHC.  In Fig.~\ref{figbetaLR}, taken again from \cite{Tello:2010am}, we show $|m_ {\nu +N}^{ee}|$ as a function of the
lightest neutrino mass.  We have already stressed in the introduction the reversed role of the
neutrino mass hierarchies.  In the case of the right-handed contribution, the normal hierarchy (NH)
prevails over the inverted (IH) in wide regions of the parameter space and for both
hierarchies new physics can win over the neutrino mass as the source of \BB. Moreover,
Fig.~\ref{figbetaLR} shows that there is no more room for cancellations, present in the individual
contributions in Fig.~\ref{figTelloPlot}. On the upper horizontal axis, we also display the lightest
of the heavy neutrinos. As one can see, the range of $m_N^{\text{lightest}}$ is easily below
100\,\GeV\ which would lead to interesting displaced vertices at LHC~\cite{Maiezza:2010ic}.

 In short, LR theory at the LHC energies makes a strong case for the neutrinoless double beta decay. In this sense, it is rather
 different from the usual simple seesaw picture, where it is neutrino mass is behind this process. This is clear for all three types
 of seesaw and holds true thus for the SU(5) theory which predicts TeV scale fermion triplet.

\section {Same sign lepton pairs at colliders}   

\begin{figure}[h!]
\begin{center}
\includegraphics[scale=0.55]{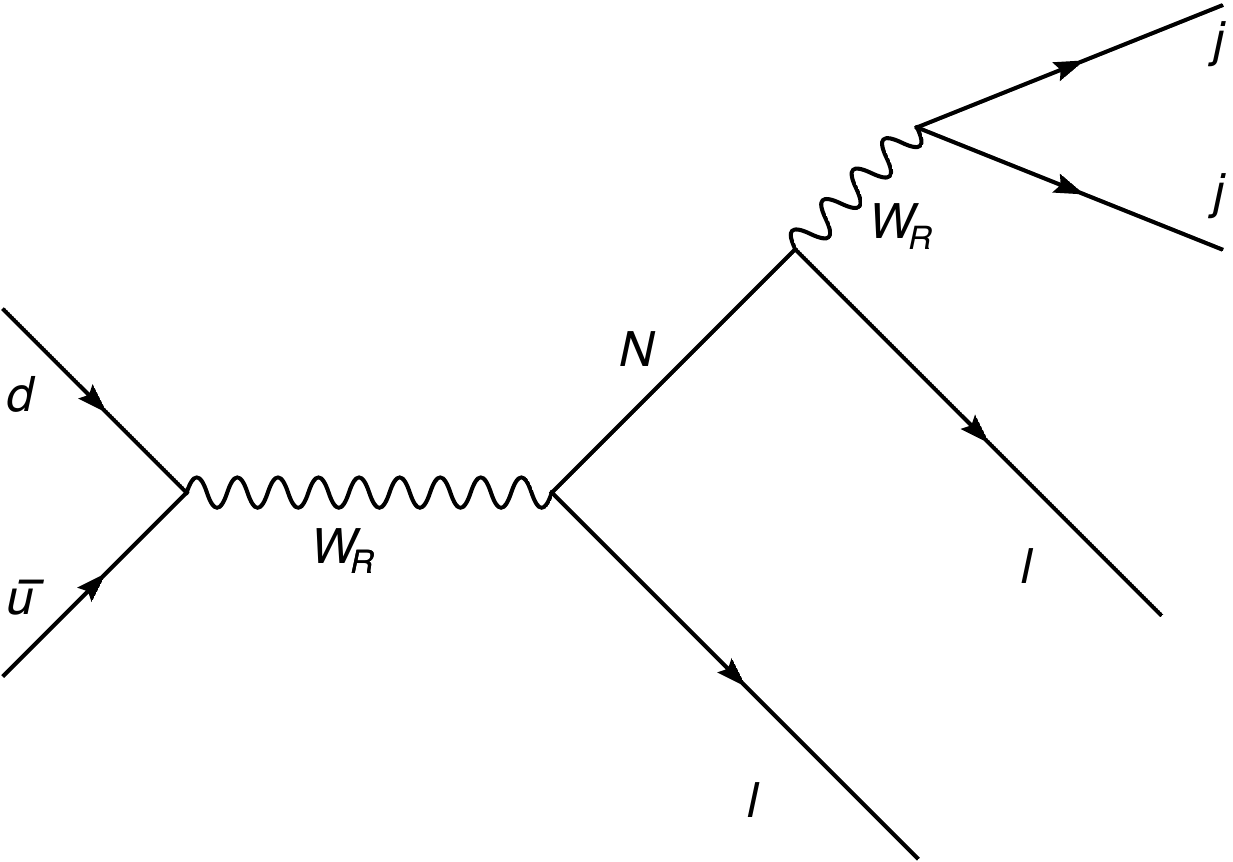} 
\caption{The production of $W_R$ and the subsequent decay into same sign leptons and two jets through
the Majorana character of the right-handed neutrino.}
\label{wr}
\end{center}
\end{figure}

The golden event for the colliders is provided by the same sign lepton pairs through a $W_R$ production, see Fig. \ref{wr}.

\begin{figure}
\begin{center}
\includegraphics[scale=0.4]{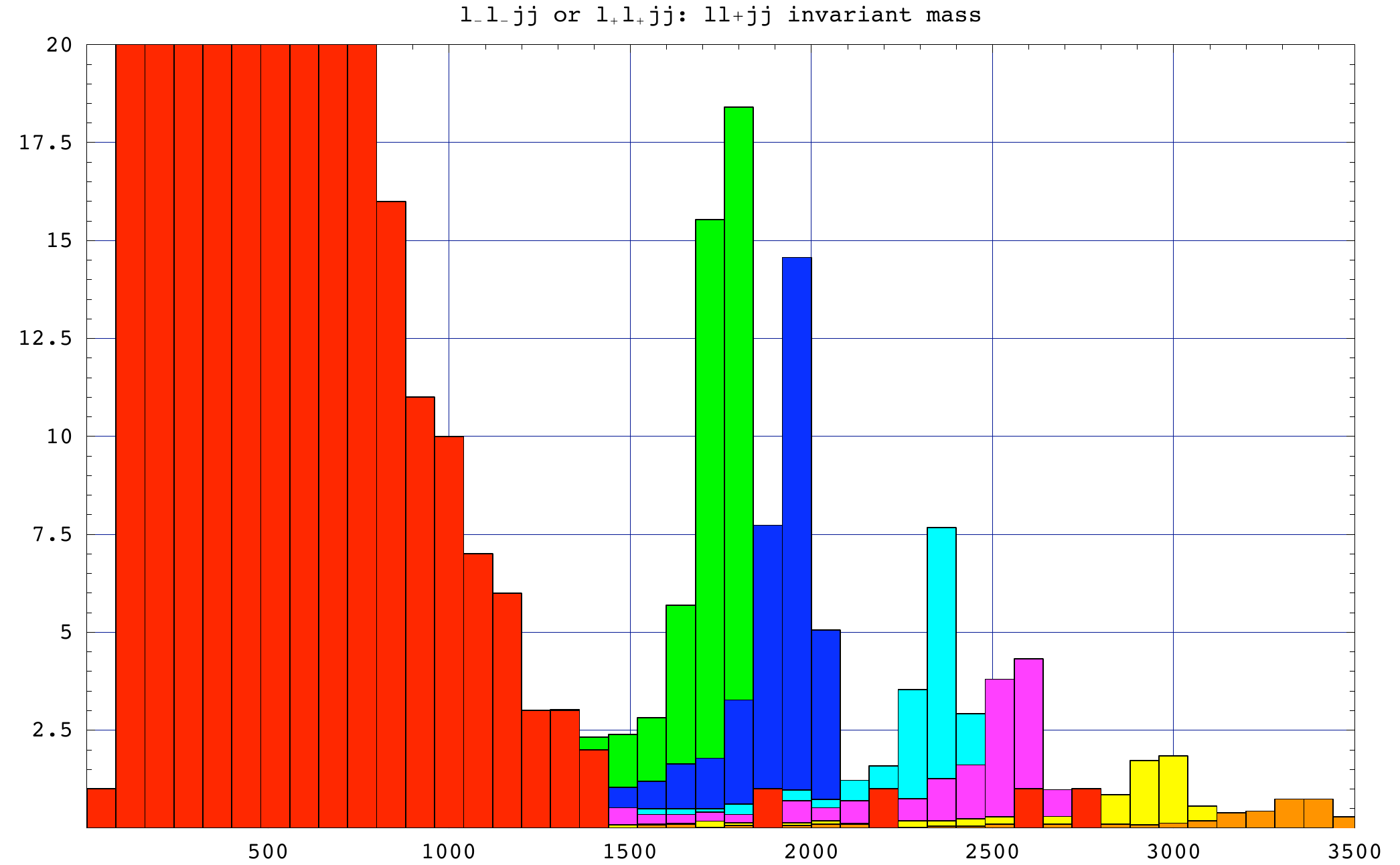} 
\caption{The expected number of events at the 14 TeV LHC as a function of energy (GeV) for ${\rm L}=8{\rm fb} ^{-1}$ (courtesy of F. Nesti). For details see \cite{Maiezza:2010ic}.}
 \label{dilep}
\end{center}
\end{figure}

Once the right-handed gauge boson is produced, it will decay into a right-handed neutrino and a charged lepton.  The 
right-handed neutrino, being a Majorana particle, decays equally often into charged leptons or anti-leptons and jets.
 In turn, one has exciting events of same sign lepton pairs and two jets, as a clear signature of lepton number violation.
This is a collider analogue of neutrinoless double beta decay, and it allows for the determination of $W_R$ mass as
shown in the Fig. \ref{dilep}.

  This offers 
\begin{list}{}{}
\item a) a direct test of parity restoration through a  discovery of $W_R$,
\item b) a direct test of lepton number violation through a Majorana nature of $\nu_R$,
\item c) determination of $W_R$  and $N$ masses, and the right-handed leptonic mixing matrix $V_R$.
With $V_R$ determined, one can make the predictions for $0\nu 2 \beta$ and LFV which we illustrated with
the type II seesaw when  $V_L = V_R^*$.
\end{list}

  A detailed study \cite{Ferrari:2000sp} concludes an easy  probe of $W_R$ mass up to 4 TeV and $\nu_R$ mass in 100 - 1000 GeV
     for integrated luminosity of 30 ${\rm fb} ^{-1}$.  
%

The flavor dependence of $V_R$ can be determined precisely through these same sign lepton pair
channels; thus, Eq.~\eqref{rightmixing} can be falsified in the near future.  Moreover, if LHC will
measure the heavy right-handed masses in the same process one could perform crucial consistency
checks of type II seesaw \cite{Tello:2010am}, such as  
\begin{equation}\label{oscdata}
\frac{m_{N _2}^2-m_{N_1}^2}{m_{N _3}^2-m_{N _1}^2}=
\frac{m_{\nu _2}^2-m_{\nu_1}^2}{m_{\nu_3}^2-m_{\nu_1}^2} \simeq \pm 0.03\,,
\end{equation}
where the right-hand side is determined by oscillation data and the $\pm $ signs corresponds to
normal/inverted hierarchy case.  Another eloquent relation among the mass scale probed in cosmology,
atmospheric neutrino oscillations and LHC was derived in \cite{Tello:2010am}
\begin{equation}\label{eqGoldenFormula}
  m_{\text{cosm}} = \sum m_{\nu_i} \simeq
  50\,\meV\times\! \frac{\sum_i m_{N_i} }{\sqrt{|m_{N_3}^2 - m_{N_2}^2|}}.
\end{equation}
To summarize, the measurement of the heavy mass spectrum can easily invalidate the model in question.

On top, the type II seesaw employed here offers another potentially interesting signature:
pair production of doubly charged Higgses  which decay into same sign lepton (anti lepton) pairs \cite{Han:2007bk}.  This can serve as
a determination of the neutrino mass matrix in the case when type I is not present or very small \cite{Kadastik:2007yd}.

Finally, a small digression. It is noteworthy that the supersymmetric version of this theory \cite{Aulakh:1998nn} predicts 
doubly charged scalars at the collider energies \cite{Aulakh:1998nn} \cite{Chacko:1997cm} \cite{Aulakh:1997fq} even
for large scale of left-right symmetry breaking. The supersymmetric version offers a rather interesting possibility~\cite{FileviezPerez:2008sx}
of getting rid of new Higgses $\Delta_{L,R}$, since the right-handed sneutrino can serve the same purpose~\cite{Hayashi:1984rd}.
 This implies that the scale of LR symmetry breaking must be at TeV, if one sticks to the usual picture of low energy
supersymmetry, and on top, one gets only one heavy right-handed neutrino at the same scale \cite{Mohapatra:1986su},
\cite{Ghosh:2010hy}, \cite{Barger:2010iv}. A careful study then reveals \cite{Ghosh:2010hy} that the remaining two right-handed neutrinos must be light, in the sub-eV region. This seems to fit with the BBN~\cite{Hamann:2010bk}, as remarked above. Even if one includes the $\Delta$ Higgs fields, a renormalizable version of the theory requires 
a non-vanishing right-handed sneutrino vev in order that electromagnetic charge invariance need not be broken~\cite{Kuchimanchi:1993jg}. Again, the LR scale would be tied to the scale of low energy supersymmetry~\cite{Kuchimanchi:1995vk} with some interesting resulting phenomenology~\cite{Chen:2010ss}.

\section{Summary and Outlook}

I discussed here an  experimental probe of Majorana neutrino mass origin, both at colliders 
through the production of the same sign di-leptons,
and through neutrinoless double beta decay.  A classical example is provided  by the $L-R$ symmetric theory that predicts
the existence of right-handed neutrinos and leads to the seesaw mechanism.
A TeV scale $L-R$ symmetry, as discussed here, would have spectacular signatures at LHC,
with a possible discovery of $W_R$ and $\nu_R$.  This offers a possibility of observing parity restoration and the Majorana nature of neutrinos. 
Furthermore, the measurements at the colliders can fix the masses and the mixings of the right-handed neutrinos, which in turn can make
predictions for the neutrinoless double beta decay and lepton flavor violation. This is the essence of our recent work  \cite{Tello:2010am}, and it
will be discussed at length in near future \cite{future}.
  
  One of the main messages that I wish to convey is that, contrary to the conventional claims in the literature, neutrinoless
  double beta decay may be dominated by new physics and not by neutrino masses. A priori, this process is not a probe
  of neutrino Majorana mass. It can even happen that the cosmological data invalidate completely this possibility if they keep
  bringing down the sum of neutrino masses and if the new experiments were to confirm a claim of this process being seen, corresponding to $m_\nu^{ee} \approx 0.4\,\eV$ \cite{KlapdorKleingrothaus:2004wj}. Actually, it was already
 argued that the two are incompatible~\cite{Fogli:2008ig}. If it were to be true, new physics would be a must. This would
 be great news for if new physics is a source of the neutrinoless double beta decay it must be at the TeV scale in order
 to provide a large enough effect.  In other words, new physics behind neutrinoless double beta decay is at the LHC reach.
  
    Furthermore, since lepton number violation is sensitive to higher scales, the LHC scale physics is likely
 to lead to observable LFV processes. A particularly exciting is the case of $\mu \to e$ conversion in nuclei if the 
 planned increased in sensibility by four to six orders of magnitude gets realized. This would open the door for measuring
 charged lepton phases, otherwise hard to measure at the LHC. 
 
  \paragraph{Comparison with SU(5).} It is clear that LR symmetry at LHC is more than exciting, and it is true that eventually neutrinoless double decay, if observed 
   in near future, could demand it to be at the TeV energy scale,  but  the theory does not predict this. As discussed in the 
   introduction, a simple SU(5) theory with an adjoint fermion field does predict the TeV seesaw mechanism ~\cite{Bajc:2006ia}. 
   What about its signatures? First and foremost, in this theory, as in the simple seesaw pictures, the neutrinoless double
   beta decay is due to neutrino mass, and if this were to be ruled out so would be this theory. Second, LFV is not generic, and
 it cannot serve as a strong constraint on the theory~\cite{Kamenik:2009cb}. It does predict LFv at colliders, but in a rather 
 distinctive manner, with four jets instead of two. However, it is a truly exciting possibility, due to the predicted low mass of the
 fermion triplet as a source of neutrino masses. It is worth to summarize some essential features of this triplet at LHC.
 
 It can be produced through gauge interactions (Drell-Yan)
\begin{eqnarray}
pp\to W^\pm +X\to T^\pm T^0 +X\nonumber\\
pp\to (Z\,{\rm or}\, \gamma)+X\to T^+T^-+X\nonumber
\end{eqnarray}

The best channel is a pair of like-sign leptons plus jets~\cite{Arhrib:2009mz}
\begin{eqnarray}
BR(T^\pm T^0\to l_i^\pm l_j^\pm +4{\rm jets})\approx\frac{1}{20}\times
\frac{|y_T^i|^2|y_T^j|^2}{(\sum_k|y_T^k|^2)^2}\nonumber
\end{eqnarray}

where $y_T^i$ are Yukawa couplings of the triplet T to the leptonic doublets.  

\begin{figure}[tb]
\begin{center}
\includegraphics[scale=0.4]{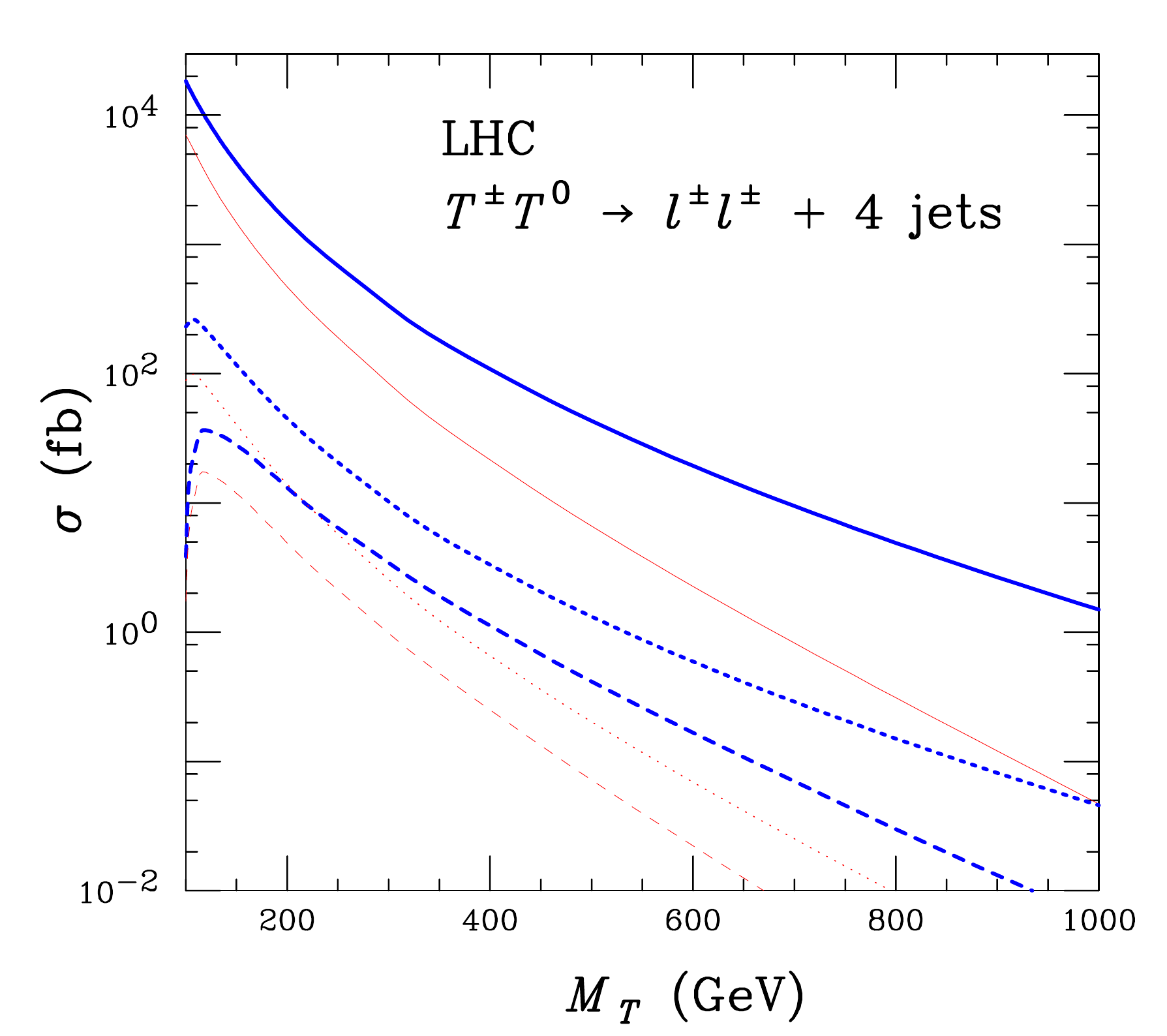}
\end{center}
\caption{
Total cross section for $p p \to T^\pm  T^0$ production and decay at the LHC 
 at $\sqrt S = 14$ TeV (thick curves) and 7 TeV (thin curves) versus the heavy lepton mass.
The solid curves (top) are for the production rate before decay or cuts.
 The dotted (middle) curves includes branching fraction of the leading channels for the case of inverse hierarchy.
 The dashed (lower) curves further include the selection cuts. For details see~\cite{Arhrib:2009mz}.
}
 \label{fig:cs_sig}
 \end{figure}

The cross sections for 
the production and subsequent decay of the triplet are sizable both at Tevatron and LHC. In figure  
~\ref{fig:cs_sig}
this is given for 
LHC at both 7 and 14 TeV center-of-mass energy.  A careful study~\cite{Arhrib:2009mz} shows that the triplets
can be searched for up to 450 (700) GeV at LHC with 14 TeV C.M. energy and 10 (100) $fb^{-1}$ luminosity.

Besides the triplet $T$, the theory contains also a singlet $S$. After the SU(2)$_L\times$U(1) symmetry breaking $\langle H\rangle=v$, 
one obtains in the usual manner the light neutrino mass matrix upon integrating out 
$S$ and $T$
\begin{equation}
\label{seesaw13}
m_\nu^{ij}=-v^2\left(\frac{y_T^iy_T^j}{M_T}+\frac{y_S^iy_S^j}{M_S}\right).
\end{equation}
 
 On the other hand,  
the triplet decay width into the k-th lepton is proportional to 
 \begin{equation}
 \Gamma_T  \propto  M_T  |y^k_T|^2 , 
\end{equation}

The same couplings $y_T^i$ contribute thus to $\nu$ mass matrix and $T$ decays, so that T decays can serve to probe the
neutrino mass matrix ~\cite{Arhrib:2009mz},~ \cite{Bajc:2006ia} and the nature of the hierarchy of neutrino masses. The main 
reason for this is the fact that the model predicts only two massive neutrinos, the lightest one effectively massless. 
Let us give an example of the inverse hierarchy for small $\theta_{13}$ (taken to be zero). 
   One finds~ \cite{Bajc:2006ia}

\begin{equation}
\frac{\rm BR_{\tau}}{\rm BR_{\mu}}=\tan^2{\theta_{23}}
\end{equation}
where $BR_{\tau}$ and $BR_{\mu}$ are branching ratios for the T decay into tau leptons and muons.

Thus
LHC can allow one to make predictions for the neutrinoless double beta decay. It is an example of a theory that uses collider data
in order to shed light on the neutrinoless double decay, but in this case it is neutrino mass does the job. The connection is not 
as direct as in the case of LR symmetric theory, but still, one can have a complementary possibility of determining neutrino  mass
hierarchy.
Although  this theory may not be 
as beautiful and as the LR symmetric one, its predictivity make it stand out among seesaw
theories of neutrino masses.  And moreover, it relates the LHC energy physics with the proton decay~\cite{Arhrib:2009mz}.

   In summary, the physics discussed in this talk offers a deep and close
 connection between high energy experiments such as LHC, and low energy ones, such as neutrinoless double beta decay
 and LFV processes. We look forward with excitement to the era ahead, and keep in mind that LHC is not only a Higgs
 hunting machine, or supersymmetry and extra dimensions one. I hope to have convinced you that it has all the potential to probe the origin of neutrino mass, the only new physics beyond the standard model observed with certainty up to date.
 
 \section{Acknowledgements}

  I wish to thank Harald Fritzsch and other organizers of the Gell-Mann Fest for an excellent conference and a warm
  hospitality.
     I am
  deeply grateful to my collaborators on the topics covered above   
   Alessio Maiezza, Alejandra Melfo, Miha Nemev\v{s}ek, Fabrizio Nesti,  
    Vladimir Tello, Francesco Vissani and Yue Zhang. 
  I acknowledge with great pleasure my old collaborations with Rabi Mohapatra on LR symmetry and seesaw, and with Wai-Yee Keung
  on lepton number violation at colliders. Thanks are due to Alejandra Melfo and Miha Nemev\v{s}ek for careful reading of this
manuscript. This work was partially supported by the EU FP6 Marie Curie Research and Training Network "UniverseNet" (MRTN-CT-2006-035863).

\end{document}